\documentclass[twocolumn]{IEEEtran}
\usepackage{amsmath}
\usepackage{amssymb}
\usepackage{amsfonts}
\usepackage{graphicx}
\usepackage{epsfig}
\usepackage{subfigure}
\usepackage{psfrag}
\usepackage{cite}
\usepackage{latexsym}
\usepackage{url}
\usepackage{color}

\begin{document}
\title{A Generic Receiver Architecture for MIMO Wireless Power Transfer with Non-Linear Energy Harvesting
\thanks{G. Ma and J. Xu are with the School of Information Engineering, Guangdong University of Technology, Guangzhou, China (e-mail: gangma.gdut@gmail.com, jiexu@gdut.edu.cn). J. Xu is the corresponding author.}
\thanks{Y. Zeng is with the School of Electrical and Information Engineering, The University of Sydney, NSW 2006, Australia (e-mail: yong.zeng@sydney.edu.au).}
\thanks{M. R. V. Moghadam is with the TransferFi Pte Ltd, Singapore. (e-mail: reza@transferfi.com).}}
\author{Ganggang Ma, {\it Student Member, IEEE}, Jie Xu, {\it Member, IEEE}, Yong Zeng, {\it Member, IEEE}, \\and Mohammad R. Vedady Moghadam, {\it Member, IEEE}\vspace{-1em}}
\maketitle

\begin{abstract}
This letter investigates a multiple-input multiple-output (MIMO) wireless power transfer (WPT) system under practical non-liner energy harvesting (EH) models. We propose a new generic energy receiver (ER) architecture consisting of $N$ receive antennas and $L$ rectifiers, for which one power splitter is inserted after each antenna to adaptively split the received radio frequency (RF) signals among the $L$ rectifiers for efficient non-linear RF-to-direct current (DC) conversion. With the proposed architecture, we maximize the total harvested DC power at the ER, by jointly optimizing the transmit energy beamforming at the energy transmitter (ET) and the power splitting ratios at the ER. Numerical results show that our proposed design by exploiting the nonlinearity of EH significantly improves the harvested DC power at the ER, as compared to  two conventional designs.
\end{abstract}


\vspace{-1em}

\newtheorem{theorem}{\underline{Theorem}}[section]
\newtheorem{lemma}{\underline{Lemma}}[section]
\newtheorem{proposition}{\underline{Proposition}}[section]
\newtheorem{remark}{\underline{Remark}}[section]
\newcommand{\mv}[1]{\mbox{\boldmath{$ #1 $}}}

\section{Introduction}\label{sec:Introduction}
Wireless power transfer (WPT) via radio frequency (RF) signals has emerged as a promising solution to provide convenient energy supply to low-power devices in Internet of things (IoT) networks \cite{Yong,Feedback,UAV}. With WPT, dedicated energy transmitters (ETs) are deployed to transmit RF signals to charge energy receivers (ERs) connected to IoT nodes. WPT has found a wide range of applications in wireless networks, such as simultaneous wireless information and power transfer (SWIPT) \cite{Rui,SplittingPaper}, wireless powered communication \cite{WPCN,DongInKim}, and wireless powered mobile edge computing \cite{Feng,Suzhi}. In order to combat the severe signal propagation loss over distance, the multi-antenna transmit energy beamforming technique has been proposed to steer RF signals towards desired directions for intended ERs, thus improving the energy transfer efficiency \cite{Feedback,Rui}.

Conventional WPT literature (see, e.g., \cite{Feedback,UAV,Rui,SplittingPaper}) mostly considers linear energy harvesting (EH) model at ERs for the convenience of analysis, i.e., the RF-to-direct current (DC) conversion efficiency at ER is assumed to be a constant for simplicity. In practice, however, the RF-to-DC conversion process at each rectifier is highly non-linear. Specifically, as the input RF power increases, the RF-to-DC conversion efficiency is first increasing and then decreasing in general, and there exists an optimal input RF power level leading to the maximum conversion efficiency. Furthermore, when the input RF power exceeds a certain threshold, the output DC power saturates (see, e.g., [10, Fig. 4]). Due to the non-linear EH model, the conventional ER architectures in the multi-antenna WPT/SWIPT literature (see, e.g., \cite{Feedback,XiongKe,non-linearModel}), cannot efficiently exploit the RF energy from the ET as the input power level may vary. For instance, in a widely-adopted ER architecture in \cite{XiongKe,non-linearModel}, the receive RF signals from multiple antennas are combined into one single rectifier for RF-to-DC conversion. If the input RF power is high in this case, then the combined RF power from multiple receive antennas may exceed the saturation power, thus limiting the overall energy conversion efficiency. In another ER architecture \cite{Feedback}, each receive antenna is connected to one dedicated rectifier for RF-to-DC conversion independently. If the received RF power is low, then each rectifier may work at a point with very low RF-to-DC conversion efficiency. To overcome such issues, it is desirable to design new ER architectures for more efficient EH by exploiting the non-linear RF-to-DC conversion, which thus motivates our study in this work.

\begin{figure}
\centering
 \epsfxsize=1\linewidth
    \includegraphics[width=8.5cm]{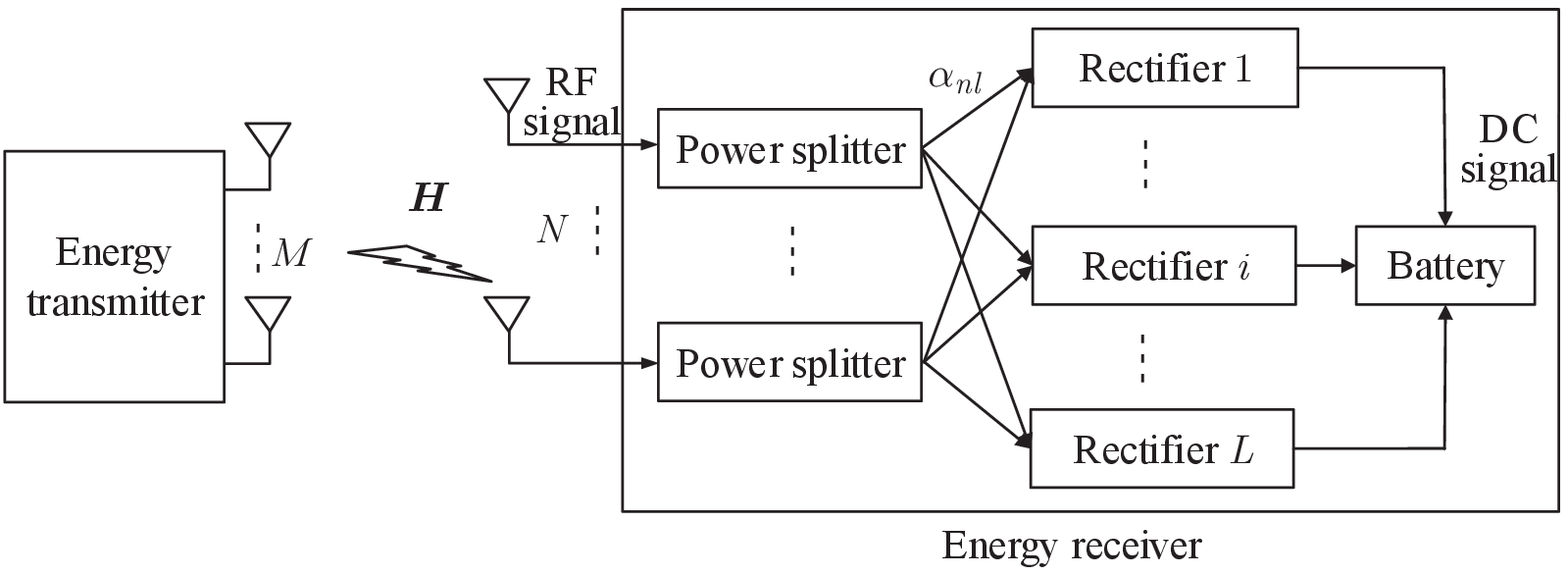}
\caption{A point-to-point MIMO WPT system with the proposed generic ER architecture consisting of $N$ receive antennas and $L$ rectifiers.} \label{fig:model}
\vspace{-2.1em}
\end{figure}
In this letter, we study a point-to-point multiple-input multiple-output (MIMO) WPT system with $M$ transmit antennas at the ET and $N$ receive antennas at the ER, as shown in Fig.~\ref{fig:model}. To analytically model the EH non-linearity with respect to input RF power, we adopt the sigmoidal function-based model in \cite{XiongKe,non-linearModel,KHuang}, where the function parameters are determined based on curve fitting to match with practical measurement results. Under this setup, we propose a novel generic ER architecture consisting of $L>1$ rectifiers, in which one power splitter is inserted after each antenna to adaptively split the received RF signals towards the $L$ rectifiers for efficient RF-to-DC conversion. Intuitively, if the received RF power level is sufficiently low, then the ER can combine all RF signals in one single rectifier; while if it is sufficiently high, then the ER can equally split them into $L$ rectifiers, thus working in the linear RF-to-DC conversion regime to maximize the WPT efficiency. Nevertheless, how to optimize the received DC power by adaptively controlling the power splitting ratios under general input power levels is unknown. Hence, in this work, we jointly optimize the transmit energy beamforming at the ET and the power splitting ratios at the ER to maximize the total harvested DC power at the ER. Although the optimization problem is non-convex and challenging to solve, we propose an efficient algorithm to obtain the global optimal solution. Numerical results show that our proposed design significantly improves the harvested DC power at the ER, as compared to the two conventional designs above.

It is worth noting that in the RF literature, the authors in [14] proposed a reconfigurable EH circuit design with multiple rectifiers (or stages), such that the number of stages can be reconfigured to accommodate variable input RF power to improve the energy conversion efficiency. However, this work considered only one single receive antenna (versus the multiple antennas in this paper), and focused on the receiver design only (versus the joint transmitter and receiver design in this paper). It is worth remarking that the sigmoidal function-based model considered in this paper is only accurate at medium and large input power regimes (e.g., larger than 0.1 mW), but inaccurate at low input power regimes (due to the variation of transmit signal distributions) \cite{BrunoJesc}. When the input power is small, there is another non-linear EH model based on a Taylor expansion of the diode's characteristics \cite{Bruno,BFwaveform}. Under such non-linearity in this regime, the design of transmit signal distributions or waveforms is crucial to further increase the harvested DC power. As our main focus is on the medium and large input power regimes, we consider the fixed transmit waveforms in this paper, and will extend to the case with adaptive waveform optimization in our future work.


\section{System Model}\label{sec:System}

As shown in Fig. \ref{fig:model}, we consider a point-to-point MIMO WPT system, where an ET equipped with $M$ antennas transmits RF power to an ER with $N$ antennas and $L$ rectifiers. With our proposed receiver architecture at the ER, $N$ power splitters are inserted between $N$ receive antennas and $L$ rectifiers, such that the received RF power at each antenna is split into $L$ portions, each for one rectifier. After that, each rectifier converts the combined RF signals from different antennas into DC signals for EH. Let $\mathcal M \triangleq\{1,..., M\}$, $\mathcal N \triangleq\{1,..., N\}$, and $\mathcal L \triangleq\{1,..., L\}$ denote the corresponding sets of antennas and rectifiers.

Let ${\mv x} \in {\mathbb C}^{M\times 1}$ denote the energy-bearing signal sent by the ET, and ${\mv h}_n \in {\mathbb C}^{M\times 1}$ denote the channel vector from the ET to the $n$-th antenna of the ER. The received RF power at the $n$-th receive antenna can be written as
\begin{align}
Q^{\rm a}_n = \mathbb E\left(|{\mv h}^H_n \mv x|^2\right) = {\mv h}^H_n {\mv X} {\mv h}_n,~n\in {\mathcal N},
\label{equa:system:1}
\end{align}
where ${\mv X} \triangleq {\mathbb E \left(\mv {xx}^H\right)}$ denotes the transmit covariance matrix at the ET, which is positive semi-definite, i.e., ${\mv X} \succeq \mv 0$. Here, $\mathbb{E}(\cdot)$ denotes the statistical expectation, $|x|$ represents the magnitude of a complex number $x$, and the superscript $H$ denotes the conjugate transpose of a matrix. Let $P_{\rm max}$ denote the maximum transmit power of the ET. Then we have ${\rm tr}({\mv X}) \le P_{\rm max}$, where $\rm{tr}(\cdot)$ denotes the trace of a square matrix. Let $\alpha_{nl}\in [0,1]$ denote the power splitting ratio from the $n$-th antenna to the $l$-th rectifier, $n\in \mathcal N$, $l\in \mathcal L$, where $\sum\limits_{l\in \mathcal L}\alpha_{nl}=1$, $\forall n\in \mathcal N$. Therefore, the combined input RF power at the $l$-th rectifier is given as
\begin{align}
Q^{\rm b}_l &= \sum\limits_{n\in \mathcal N} \alpha_{nl}Q^{\rm a}_n= \sum\limits_{n\in \mathcal N} \alpha_{nl}{\mv h}^H_n {\mv X} {\mv h}_n.
\label{equa:system:2}
\end{align}

As for the RF-to-DC energy conversion at the $l$-th rectifier, we adopt the non-liner EH model in \cite{non-linearModel}, where the output DC power at the $l$-th rectifier can be written as a function of the input RF power $Q_l^{\rm b}$ as{\footnote{Notice that this analytic non-linear EH model is sufficiently accurate at the medium to large input power regimes (e.g., larger than 0.1 mW) \cite{BrunoJesc}.}
\begin{align}
Q^{\rm DC}_l(Q_l^{\rm b}) = \frac{Q^{\rm dc}_l(Q_l^{\rm b})-Q^{\rm max}_l \Omega_l}{1-\Omega_l},
\label{equa:system:3}
\end{align}
with
\begin{align}
&\Omega_l = \frac{1}{1+e^{a_l b_l}}, \label{equa:system:4}\\
&Q^{\rm dc}_l(Q_l^{\rm b})= \frac{Q^{\rm max}_l}{1+e^{-a_l(Q^{\rm b}_l-b_l)}}.
\label{equa:system:5}
\end{align}
Here, $\Omega_l$ is a constant to ensure the zero-input/zero-output response for EH, and $Q^{\rm dc}_l(Q^{\rm b}_l)$ is the sigmoid function with respect to the input RF power $Q^{\rm b}_l$. Furthermore, $Q^{\rm max}_l$ denotes the maximum output DC power at the $l$-th rectifier when it is saturated. Also, $a_l$ and $b_l$ are two constants depending on specific circuit parameters such as the resistance, capacitance, and diode, as well as the RF waveform adopted for WPT. In practice, for any given rectifier, the values of $a_l$, $b_l$ and $Q^{\rm max}_l$ can be determined by using a standard curve fitting algorithm based on the measurement results. In this case, the total harvested DC power at the ER is expressed as
\begin{align}\label{equa:total}
Q_{\rm total}=\sum\limits_{l\in \mathcal L} Q^{\rm DC}_l(Q_l^{\rm b}) =\sum\limits_{l\in \mathcal L} \frac{Q^{\rm dc}_l(Q_l^{\rm b})-Q^{\rm max}_l \Omega_l}{1-\Omega_l}.
\end{align}

Our objective is to maximize the total harvested DC power $Q_{\rm total}$ at the ER, by jointly optimizing the energy beamforming at the ET (i.e., $\mv X$) and the adaptive power splitting ratios at the ER (i.e., $\{\alpha_{nl}\}$). From (\ref{equa:total}), it is observed that the total harvested DC power $Q_{\rm total}$ is a non-decreasing function with respect to the term $\sum_{l\in \mathcal L} \frac{Q^{\rm dc}_l(Q_l^{\rm b})}{1-\Omega_l}$. Therefore, the total harvested DC power maximization is equivalent to maximizing $\sum_{l\in \mathcal L} \frac{Q^{\rm dc}_l(Q_l^{\rm b})}{1-\Omega_l}$. By defining ${\tilde Q}^{\rm max}_l =\frac{Q^{\rm max}_l}{1-\Omega_l}$, the optimization problem of our interest is formulated as
\begin{align}
{\mathtt{(P1)}}: \max\limits_{\mv{X},~\{\alpha_{nl}\}}~ &\sum\limits_{l \in\mathcal L} ~ \frac{{\tilde Q}^{\rm max}_l}{1+e^{-a_l(\sum\limits_{n\in\mathcal N} \alpha_{nl}{\mv h}^H_n {\mv X} {\mv h}_n-b_l)}}\label{equa:optimization:1} \\
{\mathrm{s.t.}}~~~& \sum\limits_{l \in\mathcal L}\alpha_{nl}=1,~{\forall n\in \mathcal N} \label{equa:contraint:1}\\
~& 0\le \alpha_{nl} \le1,~ \forall n\in\mathcal N,~ l\in\mathcal L\label{equa:contraint:2}\\
~& {\rm tr}(\mv{X})\le P_{\rm max} \label{equa:contraint:3}\\
~& \mv{X} \succeq 0.\label{equa:contraint:4}
\end{align}
For problem (P1), we observe that the objective function is non-concave with respect to $\mv X$ and $\{\alpha_{nl}\}$. Therefore, (P1) is a non-convex optimization problem that is difficult to solve optimally in general.

\section{Proposed Solution to Problem (P1)}
In this section, we propose an efficient algorithm to solve problem (P1) optimally by first finding the optimal energy beamforming $\mv X$, and then optimizing the power splitting ratios $\{\alpha_{nl}\}$. Let $\mv{X}^\star$ and $\{\alpha_{nl}^\star\}$ denote the optimal solution to (P1).

First, we define $\mv{H}^{H}\triangleq [\mv{h}^{H}_1, ..., \mv{h}^{H}_n]$, and present the following proposition.
\begin{proposition}\label{proposition:1}
Under the non-linear model in \cite{non-linearModel},{\footnote{Note that when the diode-characteristics-based non-linear EH model in \cite{Bruno} is considered, the optimal transmit beamforming design may be coupled with the transmit sinal (see, e.g., \cite{BFwaveform}) and receiver power splitting factors. This is an interesting but challenging topic that will be left for future work.}} the optimal solution of $\mv X$ to problem (P1) is $\mv{X}^\star= P_{\rm max}\mv{v}\mv{v}^{H}$, where $\mv{v}$ denotes the eigenvector with respect to the dominant eigenvalue of $\mv{H}^{H}\mv{H}$.
\end{proposition}
\begin{IEEEproof}
See Appendix.
\end{IEEEproof}
Proposition 3.1 shows that the optimal transmit energy beamforming for the DC power maximization problem (P1) under non-linear EH model is actually identical to that in the conventional design under linear EH model \cite{Feedback,Rui}. This is due to the fact that in our design, the received RF power can be fully utilized after the adaptive power splitting towards different rectifiers.

Next, under the obtained optimal energy covariance matrix $\mv{X}^\star$, finding the power splitting ratios $\{\alpha^{\star}_{nl}\}$ for problem (P1) is equivalent to solving the following  optimization problem.
\begin{align}
{\mathtt{(P2)}}: \max\limits_{\{\alpha_{nl}\}}~ &\sum\limits_{l\in \mathcal L} ~ \frac{{\tilde Q}^{\rm max}_l}{1+e^{-a_l({\sum\limits_{n\in\mathcal N}}\alpha_{nl}{Q_n^{\rm a}}^{\star}-b_l)}}\label{equa:optimization:4} \\
{\mathrm{s.t.}}~& (\ref{equa:contraint:1})~{\text{and}}~(\ref{equa:contraint:2}),\nonumber
\end{align}
where ${Q_n^{\rm a}}^{\star}={\mv h}^H_n {\mv X^\star} {\mv h}_n$, $n\in\mathcal N$. To facilitate the presentation, we further define
\begin{align}
{\mathtt{(P3. \mv{\mu},\mv{\beta})}}: \nonumber\\
\max\limits_{\{\alpha_{nl}\}}~&\sum\limits_{l\in \mathcal L} ~ {\mu}_l\bigg[{\tilde Q}^{\rm max}_l-\beta_l\bigg(1+e^{-a_l({\sum\limits_{n\in \mathcal N}} \alpha_{nl}{Q_n^{\rm a}}^{\star}-b_l)}\bigg)\bigg] \nonumber \\
{\mathrm{s.t.}}~& (\ref{equa:contraint:1})~{\text{and}}~(\ref{equa:contraint:2}), \nonumber
\end{align}
in which $\mv{\mu} = [\mu_1,...,\mu_L]$ and  $\mv{\beta} = [\beta _1,...,\beta _L]$ are parameters for this problem. Then the following lemma holds.
\begin{lemma}\label{lemma:1}
There exist parameters $\mv{\mu}^\star=[\mu^\star_1,...,\mu^\star_L]$ and $\mv{\beta}^\star=[\beta^\star_1,...,\beta^\star_L]$ such that problem $\mathtt{(P3. \mv{\mu}^{\star},\mv{\beta}^{\star}})$ has the same optimal solution as that to problem (P2), where the optimal solution $\{\alpha_{nl}^\star\}$ satisfies the following system of equations:
\begin{align}
&{\beta}^\star_l\big(1+e^{-a_l(\sum\limits_{n\in \mathcal N} \alpha^\star_{nl}{Q_n^{\rm a}}^{\star}-b_l)}\big)-{\tilde Q}^{\rm max}_l=0, \label{equal:1}\\
&{\mu}^\star_l\big(1+e^{-a_l(\sum\limits_{n\in \mathcal N} \alpha^{\star}_{nl}{Q_n^{\rm a}}^{\star}-b_l)}\big)-1=0. \label{equal:2}
\end{align}
\end{lemma}
\begin{IEEEproof}
The proof follows directly from \cite{DPRAlgorithm} and thus is omitted for brevity.
\end{IEEEproof}

Based on Lemma \ref{lemma:1}, we solve problem (P2) by first solving problem $\mathtt{(P3. \mv{\mu},\mv{\beta}})$ under any given $\mv\mu$ and $\mv\beta$, and then finding the optimal $\mv \mu^\star$ and $\mv \beta^\star$ based on (\ref{equal:1}) and (\ref{equal:2}).

First, under any  $\mv\mu$ and $\mv\beta$, it is observed that the objective function of $\mathtt{(P3. \mv{\mu},\mv{\beta}})$ is concave, and all the constraints are linear. As a result, problem $\mathtt{(P3. \mv{\mu},\mv{\beta}})$ is convex and thus can be solved optimally via standard convex optimization techniques such as the interior point method \cite{Boyd}.

Next, we find the optimal $\mv\mu^\star$ and $\mv\beta^\star$ that satisfy the set of equations in (\ref{equal:1}) and (\ref{equal:2}) by iteratively updating them as follows.

Define $\psi_l(\mv\mu, \mv\beta)=\beta_l\big(1+e^{-a_l(\sum\limits_{n\in \mathcal N} \alpha_{nl}{Q_n^{\rm a}}^{\star}-b_l)}\big)-{\tilde Q}^{\rm max}_l$ and $\psi_{L+l}(\mv\mu, \mv\beta)=\mu_l\big(1+e^{-a_l(\sum\limits_{n\in \mathcal N} \alpha_{nl}{Q_n^{\rm a}}^{\star}-b_l)}\big)-1$, $l \in \mathcal L$. Based on \cite{DPRAlgorithm}, we find the optimal $\mv\mu^\star$ and $\mv\beta^\star$ such that  $\mv\psi(\mv\mu, \mv\beta)=[\psi_1,...,\psi_{2L}]^T=\mv0$, by using the modified Newton method in an iterative manner as follows. In the $k$-th iteration, we define
\begin{align}
\mv{p}^{(k)}=-[\mv\psi^{'}(\mv\mu^{(k)}, \mv\beta^{(k)})]^{-1}\mv\psi(\mv\mu^{(k)}, \mv\beta^{(k)}),
\label{equa:iteration:1}
\end{align}
where $\mv\psi^{'}(\mv\mu^{(k)}, \mv\beta^{(k)})$ is the Jacobian matrix of $\mv\psi(\mv\mu^{(k)}, \mv\beta^{(k)})$. Then, we update the parameters $\mv\mu$ and $\mv \beta$ as
\begin{align}
\mv\mu^{(k+1)}=\mv{\mu}^{(k)}+\lambda^{(k)}\mv{p}^{(k)}_{L+1:2L},~~ \mv{\beta}^{(k+1)}=\mv{\beta}^{(k)}+\lambda^{(k)}\mv{p}^{(k)}_{1:L},
\label{equa:iteration:2}
\end{align}
where $\lambda^{(k)}$ is the largest value of $\lambda \in (0,1]$ that satisfies
\begin{align}
&\Vert{\mv\psi(\mv\mu^{(k)}+\lambda\mv{p}^{(k)}_{L+1:2L}, \mv\beta^{(k)}+\lambda\mv{p}^{(k)}_{1:L})}\Vert \nonumber\\ &\le(1-\epsilon\lambda)\Vert{\mv\psi(\mv\mu^{(k)}, \mv\beta^{(k)})}\Vert,
\label{equa:iteration:3}
\end{align}
with $\epsilon\in (0,1)$. The detailed procedure for solving problem (P2) is summarized as Algorithm 1 in Table I. Notice that it has been shown in \cite{DPRAlgorithm} that the modified Newton method converges to a unique solution $(\mv\mu^{\star}, \mv\beta^{\star})$ of $\mv\psi(\mv\mu, \mv\beta)$ with a liner rate for any starting point, while the rate in the neighborhood of the solution is quadratic. Therefore, Algorithm 1 converges to the global optimal solution to problem (P2).

\begin{table}[t]
\begin{center}
\caption{Algorithm 1 for Solving Problem (P2)}
\end{center}
\hrule
\begin{itemize}
\item[1:] {\bf Initialize} the maximum tolerance $\Delta$, the maximum number of iterations $K$, iteration index $k=0$, $\mv\mu^{(k)}$, and $\mv\beta^{(k)}$. Choose $\lambda \in (0,1]$ and $\epsilon\in (0,1)$.
\item[2:] {\bf Repeat:}
\item[3:] \ \ \ Solve problem $\mathtt{(P3. \mv{\mu},\mv{\beta}})$ to find the optimal power splitting ratios $\{\alpha^{(k)}_{nl}\}$ under given $\mv\mu^{(k)}$ and $\mv\beta^{(k)}$.
\item[4:] \ \ \ {\bf If} $\|\mv\psi(\mv\mu, \mv\beta)\| \le\Delta$,

\ \ \ \ {\bf then} $\{\alpha^{(k)}_{nl}\}$ is the optimal solution to problem (P2).

\ \ \ \ {\bf Return} $\alpha^{\star}_{nl}=\alpha^{(k)}_{nl}$, $n\in \mathcal N$, $l\in \mathcal L$  and {\bf stop} this algorithm.
\item[5:]\ \ \ {\bf Else}
\item[6:] \ \ \ \ \ Update $\mv\mu^{(k+1)}$ and $\mv\beta^{(k+1)}$ according to (\ref{equa:iteration:1})-(\ref{equa:iteration:3}), and $k=k+1$.
\item[7:] \ \ \ {\bf End if}
\item[8:]{\bf Until} $k\ge K$.
\end{itemize}
\hrule \vspace{-2em}
\end{table}

By combining the solution of $\mv{X}^\star$ in Proposition \ref{proposition:1} and Algorithm 1 for obtaining $\{\alpha_{nl}^\star\}$, the global optimal solution to problem (P1) is finally obtained.

\vspace{-0.4em}
\section{Numerical Results}
\begin{figure*}
\begin{minipage}[t]{0.32\linewidth}
\centering
\includegraphics[width=4.5cm]{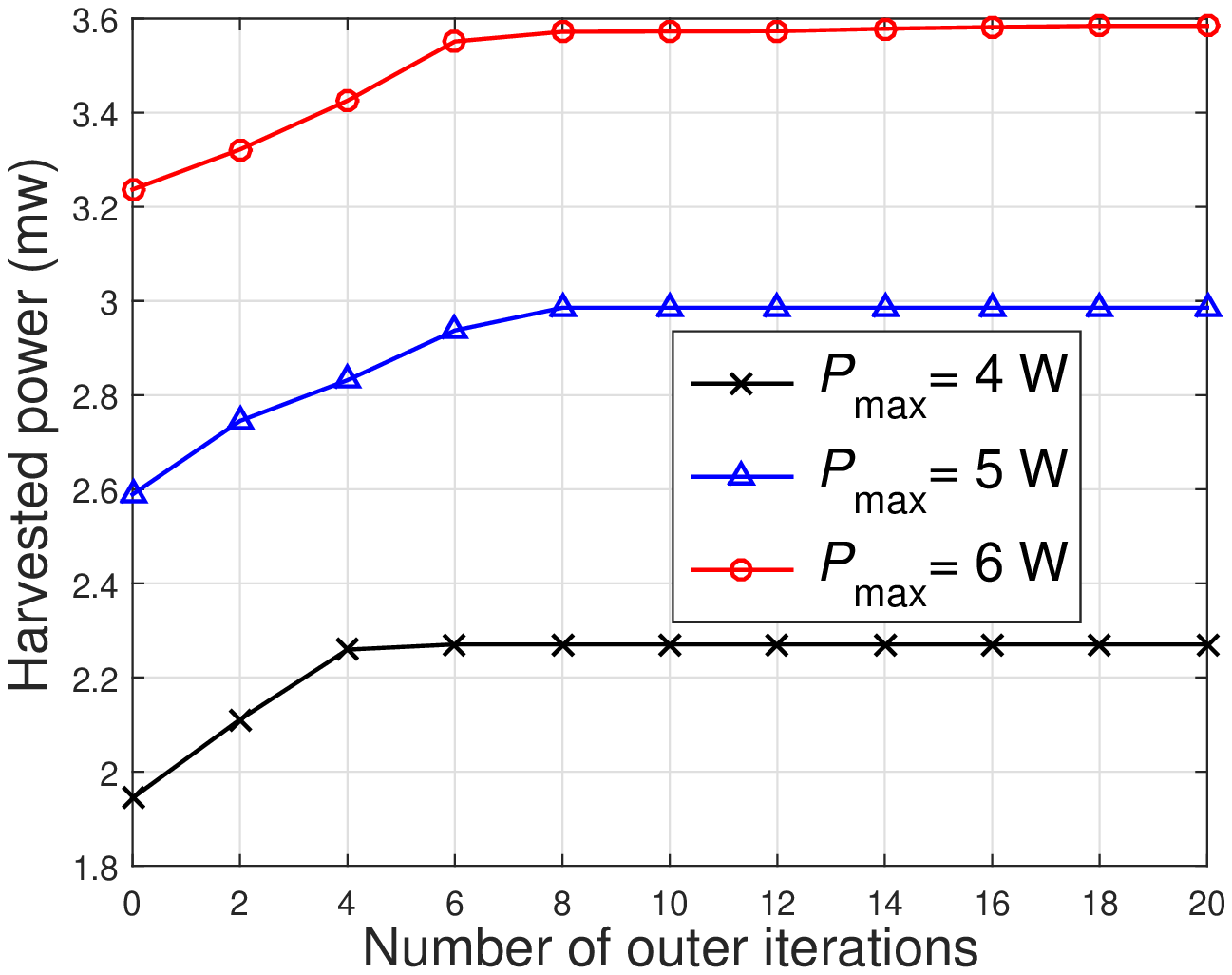}
\caption{Convergence behavior of Algorithm 1.}
\label{fig:convergence}
\end{minipage}%
\begin{minipage}[t]{0.05\linewidth}
\end{minipage}
\begin{minipage}[t]{0.32\linewidth}
\centering
\includegraphics[width=4.5cm]{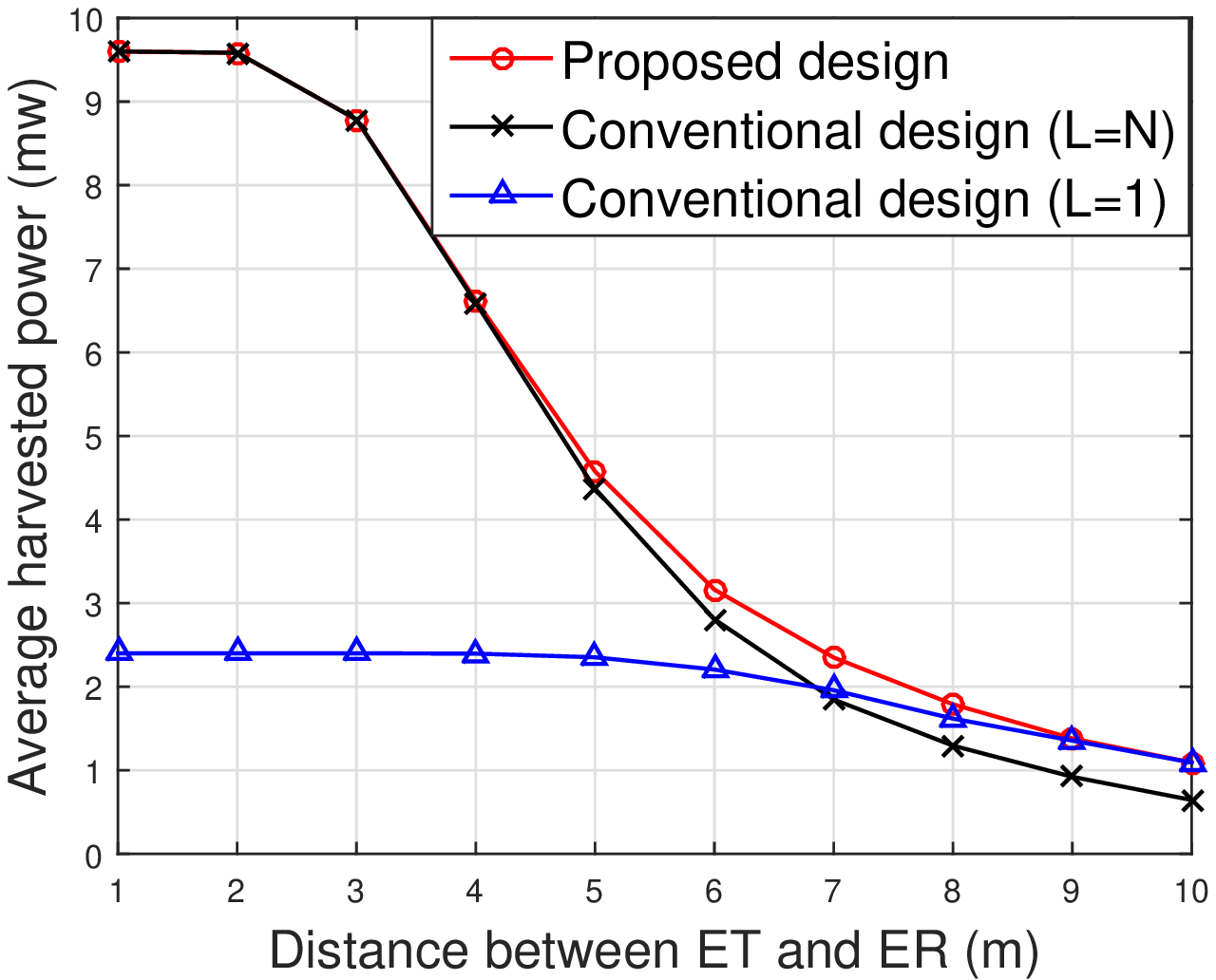}
\caption{The average harvested DC power at the ER versus the distance between the ET and the ER with the transmit power $P_{\rm max}=5$ W.}
\label{fig:power_d}
\end{minipage}
\begin{minipage}[t]{0.05\linewidth}
\end{minipage}
\begin{minipage}[t]{0.32\linewidth}
\centering
\includegraphics[width=4.5cm]{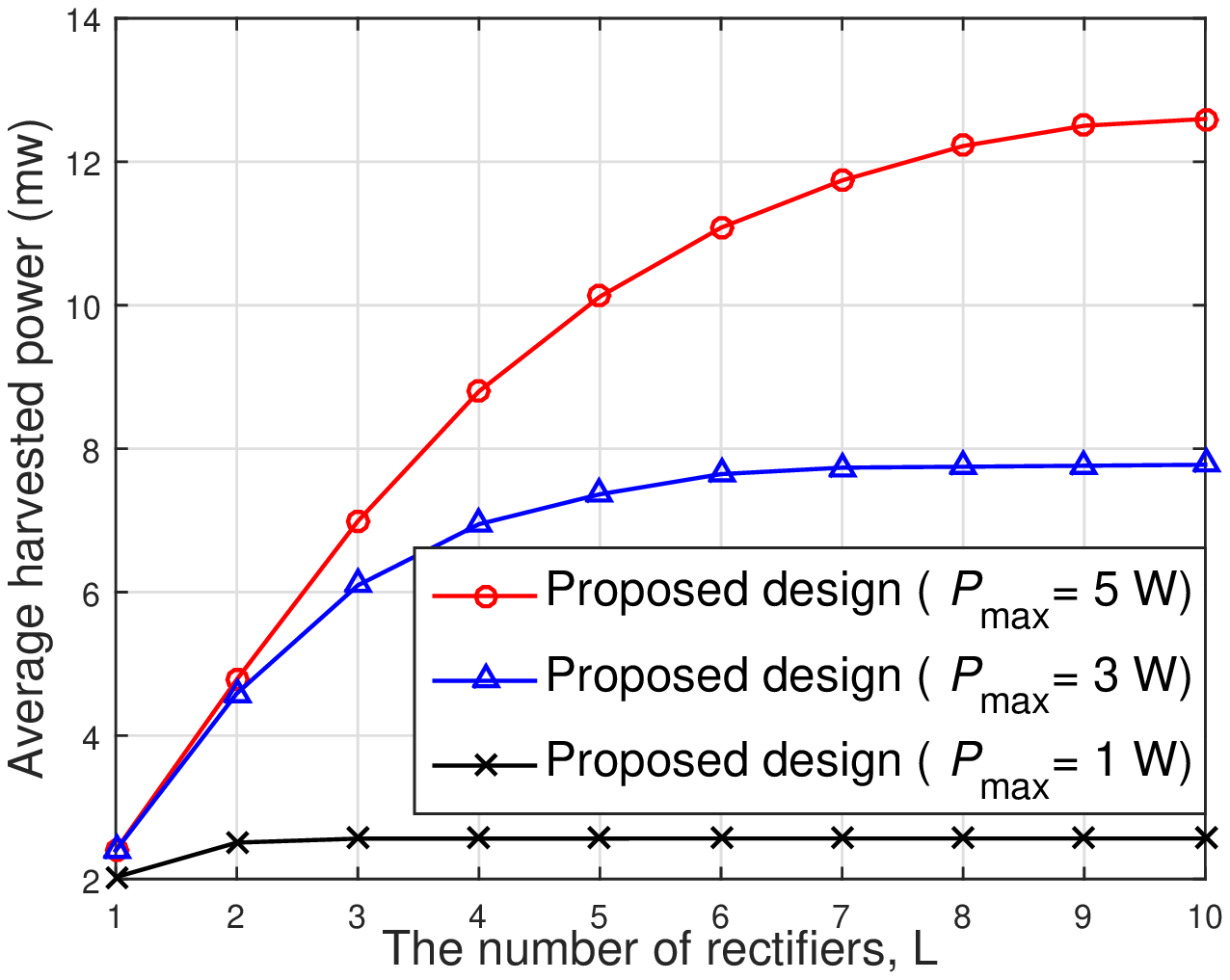}
\caption{The total harvested DC power versus the number of rectifiers $L$ with $d=3$ m.}
\label{fig:power_L}
\end{minipage}\vspace{-1.5em}
\end{figure*}
In this section, we provide numerical results to validate the performance of our proposed design, as compared to two conventional designs without power splitting as follows.

(1) {\it Conventional design with one single rectifier ($L=1$)}: In this case, the ER only has one rectifier, and the optimal transmit beamforming for the harvested DC power maximization problem is $\mv{X}^\star= P_{\rm max}\mv{v}\mv{v}^{H}$ \cite{Feedback,Rui}.

(2) {\it Conventional design with one rectifier for each antenna ($L=N$)}: At the ER, each receive antenna connects to one rectifier, with $\mathcal L = \mathcal N$. Accordingly, the harvested DC power maximization problem over $\mv X$ corresponds to maximizing $\sum\limits_{l \in\mathcal L} ~ \frac{{\tilde Q}^{\rm max}_l}{1+e^{-a_l({\mv h}^H_l {\mv X} {\mv h}_l-b_l)}}$ and subject to constraints (\ref{equa:contraint:3}) and (\ref{equa:contraint:4}).
It is observed that the objective function is non-convex, due to its nonlinear sum-of-ratios form. Despite this fact, this problem can be solved by using the modified Newton method similarly as in Algorithm 1, for which the details are omitted for brevity.

In the simulation, we consider the Rician fading channel from the ET to the ER, where the Rician factor is set as 3 dB, and the average channel power gain is $g = \zeta_0d^{-3}$. Here, $d$ is the distance between the ET and the ER and $\zeta_0=-30$~dB denotes the channel power gain at a reference distance of $d_0 = 1$ meter (m). For the non-linear EH model, we set $a_l=150$, $b_l=0.014$, and $Q_l^{\rm max} = 2.4$ mW, $\forall l\in \mathcal L$, which are obtained by curve fitting based on the measurement data \cite{rectifier}.

Fig. \ref{fig:convergence} shows the convergence behavior of Algorithm 1 in terms of the number of outer iterations under one random channel realization. It is observed that the proposed Algorithm 1 converges with a few outer iterations. This validates the effectiveness of our proposed design.


Fig. \ref{fig:power_d} shows the average harvested DC power versus the distance $d$ between the ET and ER. It is observed that our design performs best at all distance regimes, and the performance gain is more substantial when $5\ {\text m}\le d \le 8\ {\text m}$. When the distance is short with $d\le 4$ m, our proposed design is observed to achieve similar harvested DC power as the conventional design with $L = N$. This is due to the fact that in this case, the input RF power is very large, such that more rectifiers are desirable at the ER for avoid the saturation. When the distance is long with $d > 8$~m, it is observed that our proposed design achieves similar performance as the conventional design with $L=1$, which is expected since when the input power is very small, all received RF signals should be combined into one rectifier.

Fig. \ref{fig:power_L} shows the total DC harvested power at the ER versus the number of rectifiers $L$ with the distance $d=3$ m. It is observed that in the high transmit power case (e.g., $P_{\rm max}=$ 3 W or 5 W), using more rectifiers can significantly improve the harvested DC power at the ER.

\vspace{-0.5em}
\section{Conclusion}
In this letter, we propose a generic ER architecture for MIMO WPT by exploiting the non-linear EH model, where adaptive power splitters are inserted between receive antennas and rectifiers at the ER for more efficient RF-to-DC conversion. Under the proposed architecture, we jointly optimize the transmit energy beamforming at the ET and the adaptive power splitting at the ER to maximize the ER's harvested DC power. Numerical results show that our proposed design can significantly improve the harvested DC power at the ER, as compared to conventional alternatives. How to design the transmit energy beamforming, power splitting factors, jointly with the transmit waveforms for wider input RF power regimes is an interesting problem, which will be investigated in our future work.

\vspace{-0.5em}
\appendix[Proof of Proposition \ref{proposition:1}]
By introducing auxiliary variables $\mv{\varphi}=[\varphi_1, \varphi_2,..., \varphi_L]$, problem $\mathtt{(P1)}$ can be expressed as
\begin{align}
{\mathtt{(P4)}}: \max\limits_{\mv{X},~\{\alpha_{nl}\},~\{\varphi_l\}}~ &\sum\limits_{l \in\mathcal L} ~ \frac{{\tilde Q}^{\rm max}_l}{1+e^{-a_l(\varphi_l-b_l)}}\nonumber \\
{\mathrm{s.t.}}~~~& \varphi_l \le  \sum\limits_{n\in \mathcal N} \alpha_{nl}{\mv h}^H_n {\mv X} {\mv h}_n,~\forall l\in\mathcal L \label{equa:slack:1}\\
~& (\ref{equa:contraint:1}),~(\ref{equa:contraint:2}),~(\ref{equa:contraint:3}),~{\text{and}}~ (\ref{equa:contraint:4}).\nonumber
\end{align}
First, let $\mathcal X_1$ denote the feasible set of $\mv{X}$ and $\{\varphi_l\}$ specified by  constraints (\ref{equa:contraint:1}), (\ref{equa:contraint:2}), and (\ref{equa:slack:1}), and $\mathcal X_2$ denote the feasible set of $\mv X$ and $\{\varphi_l\}$ characterized by the following inequality:
\begin{align}\label{equa:proof}
\sum\limits_{l \in\mathcal L}\varphi_l \le \sum\limits_{n\in \mathcal N} {\mv h}^H_n {\mv X} {\mv h}_n.
\end{align}
Now, we show that $\mathcal X_1 = \mathcal X_2$. On one hand, by combining (8), (9), and (19), we have $\sum_{l \in\mathcal L}\varphi_l \le \sum_{n\in \mathcal N} {\mv h}^H_n {\mv X} {\mv h}_n$, and thus we have $\mathcal X_1 \subseteq \mathcal X_2$. On the other hand, for any $\mv{X}$ and $\{\varphi_l\}$ satisfying (20), we can always find a set of $\{\alpha_{nl}\}$ such that $\mv{X}$, $\{\varphi_l\}$, and $\{\alpha_{nl}\}$ satisfy (8), (9), and (19). As a result, we have  $\mathcal X_1 \supseteq \mathcal X_2$ as well. Therefore, the optimal solution of $\mv X$ and $\{\varphi_l\}$ to problem (P4) and thus (P1) is equivalent to that to the following problem:
\begin{align}
{\mathtt{(P5)}}: \max\limits_{\mv{X},~\{\varphi_l\}}~ &\sum\limits_{l \in\mathcal L} ~ \frac{{\tilde Q}^{\rm max}_l}{1+e^{-a_l(\varphi_l-b_l)}} \nonumber\\
{\mathrm{s.t.}}~~~& (\ref{equa:contraint:3}),~(\ref{equa:contraint:4}),~{\text{and}}~(\ref{equa:proof}). \nonumber
\end{align}

The maximum value of problem (P5) can be achieved only when constraint (\ref{equa:proof}) holds with the strict equality, since otherwise, $\{\varphi_l\}$ can be further increased to improve the objective value. Therefore, at the optimality of (P5), the term $ \sum_{n\in \mathcal N} {\mv h}^H_n {\mv X} {\mv h}_n$ must be maximized. Thus, the optimization over $\mv X$ in (P5) or equivalently (P1) is identical to maximizing $\sum_{n\in \mathcal N}~{\mv h}^H_n {\mv X} {\mv h}_n= {\rm tr}(\mv{H}^{H}\mv{H}\mv{X})$ under constraints (\ref{equa:contraint:3}) and (\ref{equa:contraint:4}).
It is easy to verify the optimal solution to this problem and thus (P1) is $\mv{X}^\star= P_{\rm max}\mv{v}\mv{v}^{H}$. As a result, this proposition is finally proved.

\end{document}